\documentclass[10pt, conference]{IEEEtran}
\IEEEoverridecommandlockouts
\usepackage[noadjust]{cite}
\usepackage{amsmath, amssymb, amsfonts}
\usepackage{algorithmic}
\usepackage{graphicx}
\usepackage{textcomp}
\usepackage{xcolor}

\def\BibTeX{{\rm B\kern-.05em{\sc i\kern-.025em b}\kern-.08em
    T\kern-.1667em\lower.7ex\hbox{E}\kern-.125emX}}

\begin{document}

\title{Predictive Demodulation for\\Chaotic Communications
\thanks{This work is part of the project IRENE (PID2020-115323RB-C31), funded by MCIN/AEI/10.13039/501100011033 and supported by the Catalan government through the project SGR-Cat 2021-01207.}
}

\author{\IEEEauthorblockN{Marc Martinez-Gost\IEEEauthorrefmark{1}\IEEEauthorrefmark{2}, Ana Pérez-Neira\IEEEauthorrefmark{1}\IEEEauthorrefmark{2}\IEEEauthorrefmark{3}, Miguel Ángel Lagunas\IEEEauthorrefmark{2}}
\IEEEauthorblockA{
\IEEEauthorrefmark{1}Centre Tecnològic de Telecomunicacions de Catalunya, Spain\\
\IEEEauthorrefmark{2}Dept. of Signal Theory and Communications, Universitat Politècnica de Catalunya, Spain\\
\IEEEauthorrefmark{3}ICREA Acadèmia, Spain\\
\{mmartinez, aperez, malagunas\}@cttc.es
}}

\maketitle
\begin{abstract}
Chaotic signals offer promising characteristics for wireless communications due to their wideband nature, low cross-correlation, and sensitivity to initial conditions. Although classical chaotic modulation schemes like Chaos Shift Keying (CSK) can theoretically match the performance of traditional modulation techniques (i.e., bit error rate), practical challenges, such as the difficulty in generating accurate signal replicas at the receiver, limit their effectiveness. Besides, chaotic signals are often considered unpredictable despite their deterministic nature. In this paper, we challenge this view by introducing a novel modulation scheme for chaotic communications that leverages the deterministic behavior of chaotic signals. The proposed approach eliminates the need for synchronized replicas of transmitted waveforms at the receiver. Moreover, to enhance noise robustness, we employ M-ary Frequency Shift Keying (FSK) modulation on the chaotic samples. Experimental results show that the proposed scheme significantly outperforms CSK when perfect replicas are unavailable, with the best performance achieved for low-order modulations, and resulting in minimal delay increase.
\end{abstract}

\section{Introduction}
Signals generated by chaotic systems offer promising applications for wireless communications \cite{oppenheim1992chaos}. Chaotic signals are non-periodic, exhibit a flat power spectral density and have low cross-correlation, making them resemble random-like signals, though they are completely deterministic. Their wideband nature make them suitable for implementing spread spectrum techniques. While chaotic codes cannot outperform code division multiple access (CDMA) codes in an additive white Gaussian noise (AWGN) environment, they beat Gold and m-sequences in multipath environments due to their low correlation properties \cite{martoyo2010chaos4cdma}. 
Additionally, even slight changes in initial conditions produce uncorrelated chaotic signals, theoretically enabling an infinite number of users, an advantage no other code type offers.
Given these attributes, revisiting chaotic communications for future wireless networks is crucial, particularly for their potential to offer high spectral efficiency.

Chaos shift keying (CSK) is a classical modulation for chaotic signals, where one of two chaotic sequences is chosen based on the bit value. At the receiver, the signal is correlated with copies of both chaotic sequences (i.e., replicas) and compared to determine the bit value. While CSK can achieve the bit error rate (BER) of binary phase shift keying (BPSK) \cite{lawrence2003exactber}, there are several obstacles that hinder a realistic implementation: since chaotic signals are generated by electronic circuits or lasers, it is complicated to obtain replicas of the transmitted signals at the receiver \cite{anthony2016laser}.
The extreme sensitivity of chaotic systems to initial conditions causes low correlation and, as a result, the practical performance of chaotic communications often differs significantly from that of BPSK.

To address these challenges, differential CSK (DCSK) prevents the need for a reference signal by transmitting the reference in the first half of the signal, while the second half modulates the information \cite{tse2003chaos}. This allows correlation to be computed using only the transmitted signal. However, this approach halves the data rate and raises security concerns, as the information can be easily demodulated by any receiver. 

Various other modulation schemes have been designed to address these issues, offering different trade-offs \cite{Zayegh2017modulations}. Despite these advancements, most still rely on the synchronization of reference signals at the receiver  \cite{williams2001sync}, 
\cite{kolumban1998_modulations}.

In a different fashion, a chaotic system can be decomposed into a drive-response, where the response synchronizes automatically with the output of the drive \cite{cuomo1992spread}. However, up to date, there are no chaotic synchronization techniques that can regenerate the dynamics in an AWGN environment \cite{kolumban2000sync}.

While many works claim that chaotic signals are not predictable although being deterministic \cite{oppenheim1998nonpredictable}, in this work we show the opposite. Particularly, we present a demodulation technique that relies on estimating the succeeding chaotic sample, given the current one. When the receiver knows the system dynamics of CSK, this is, what maps are used at the transmitter to generate the chaotic signals, there is not need to have replicas of the transmitted waveforms at the receiver side. Furthermore, we implement a digital $M$-ary frequency shift keying (FSK) modulation at the transmitter side, which makes the chaotic signal more robust to noise. Finally, we experimentally show that the proposed scheme outperforms traditional CSK when the receiver has imperfect replicas.

\section{System Model}
We consider a single transmitter with a stream of bits, $b_k,~k=1,\cdots,K$, to be transmitted over a communication channel. In the following we drop index $k$ and develop the scheme for a generic bit $b$. We propose a \textit{chaotic switching}, in which each bit value is assigned to different chaotic maps:
\begin{equation}
f(\cdot) =
\begin{cases}
f_0(\cdot) &
\text{if $b=0$}\\
f_1(\cdot) &
\text{if $b=1$}
\end{cases}
\end{equation}
where $f_0(\cdot)$ and $f_1(\cdot)$ are two different chaotic maps. 
For the sake of simplicity, we constrain the analysis to real, discrete and one-dimensional chaotic signals. Furthermore, we consider chaotic maps of embedding dimensions 1, this is, the chaotic sample at discrete time $n$, $x_n$, is completely determined by the previous sample $x_{n-1}$. 
Note that more complex mappings from bits to chaotic signals can be performed, such as assigning 2 bits to 4 chaotic maps. Nonetheless, we leave the analysis of these schemes for future work.

At the transmitter side, a chaotic signal $(x_0,x_1,\dots,x_{N-1})$ of length $N$ is generated according to the bit $b$ as
\begin{equation}
x_{n+1} = f_b(x_{n}),\quad b=\{0,1\},
\label{eq: chaotic_map_definition}
\end{equation}
where the first sample, $x_0$, known as the initial condition, can be chosen randomly. This is the first time where the proposed scheme deviates from conventional chaotic communications. Due to the high sensitivity of chaotic signals systems to initial conditions, slight variations on the initial condition $x_0$ make it unfeasible to obtain a replica of the chaotic signals at the receiver side. Therefore, correlations cannot be computed because the transmitted signal and the replica become uncorrelated.

The chaotic signal is modulated before transmission over a wireless communication channel. We define $\mathcal{T}:\mathbb{R}^n\rightarrow \mathbb{R}^m$ as the modulation function, which generates a signal
\begin{equation}
(y_0,y_1,\dots,y_{M-1}) =
\mathcal{T}(x_0,x_1,\dots,x_{N-1}),
\end{equation}
and $M\geq N$.
In practice, most of the literature assume that the transmitted waveform is the chaotic signal, which reduces to $y_m=x_n$ and $M=N$ \cite{tse2003chaos}.

The modulated signal is transmitted over an AWGN channel:
\begin{equation}
r_m = y_m + w_m,
~~
m=0,\dots,M-1,
\label{eq: awgn_channel}
\end{equation}
where $w_m\sim\mathcal{N}(0,N_o)$ is a Gaussian sample distributed, and $N_o$ is the noise power spectral density. We leave the consideration of fading as well as more complex channels for future work, since the development of chaotic communications is still in its infancy and noise is still a problem to study.

At the receiver side, the received signal is demodulated as 
\begin{equation}
(\tilde{x}_0,\tilde{x}_1,\dots,\tilde{x}_{N-1}) =
\mathcal{T}^{-1}(r_0,r_1,\dots,r_{M-1}),
\label{eq: received_chaotic}
\end{equation}
where $\mathcal{T}^{-1}:\mathbb{R}^m\rightarrow \mathbb{R}^n$ is the demodulation function used to recover an estimate of the chaotic signal.

\section{Predictive demodulation}
\subsection{Noiseless channel}
\label{sec: noiseless_channel}
For the sake of simplicity, assume that $b=0$. The receiver has no replicas of the transmitted chaotic signals, but knows the dynamics, this is, the two chaotic maps that have been used to generate the chaotic signals. Since chaotic signals are deterministic, the maps are used to estimate the succeeding sample for every received chaotic samples:
\begin{equation}
\hat{x}_{n+1}^{(i)} =
f_i(\tilde{x}_n),~~
n=0,\dots,N-2,~
i=\{0,1\}.
\label{eq:prediction}
\end{equation}
Note that there is no need to estimate more than one sample ahead because the dynamics of the chaotic maps are completely determined by the preceding sample, as defined in \eqref{eq: chaotic_map_definition}. 

When the channel is noiseless and there is no degradation due to the modulation or demodulation, the received chaotic samples are exactly the transmitted ones (i.e., $\tilde{x}_n=x_n,~\forall n$). Thus, the predicted samples in \eqref{eq:prediction} are exact for the matched chaotic map, but different for the other predictor. This is,
\begin{equation}
\hat{x}_{n+1}^{(0)}=x_{n+1}\quad
\text{and}\quad
\hat{x}_{n+1}^{(1)}\neq x_{n+1}
\end{equation}
Then, to estimate the underlying chaotic map given a received chaotic sequence $(\tilde{x}_0,\tilde{x}_1,\dots,\tilde{x}_{N-1})$, the receiver predicts the next sample with \eqref{eq:prediction} for the whole sequence and both maps. The resulting sequences are
\begin{equation}
(\hat{x}_{1}^{(0)},
\hat{x}_{2}^{(0)},\dots,
\hat{x}_{N-1}^{(0)})\quad
\text{and}\quad
(\hat{x}_{1}^{(1)},
\hat{x}_{2}^{(1)},\dots,
\hat{x}_{N-1}^{(1)}),
\label{eq:predicted_sequences}
\end{equation}
for which the receiver computes the distance with respect to the received chaotic signal in \eqref{eq: received_chaotic} as
\begin{equation}
d_i = \frac{1}{N-1}\sum_{n=1}^{N-1}\left(
\tilde{x}_n - \hat{x}_n^{(i)}
\right)^2,~~
i=\{0,1\}
\label{eq: distance_metric}
\end{equation}
Note that $d_0=0$ because the channel is noiseless. Besides, in \eqref{eq: distance_metric} we use the mean squared error (MSE) as the distance function, although other metrics can be used. Finally, both distances are compared to estimate the transmitted bit $\hat{b}$ as
\begin{equation}
\hat{b} =
\begin{cases}
0 &
\text{if $d_0<d_1$}\\
1 &
\text{if $d_0\geq d_1$}
\end{cases}
\end{equation}

\begin{figure*}[t]
    \centering
    \includegraphics[width=0.9\textwidth]{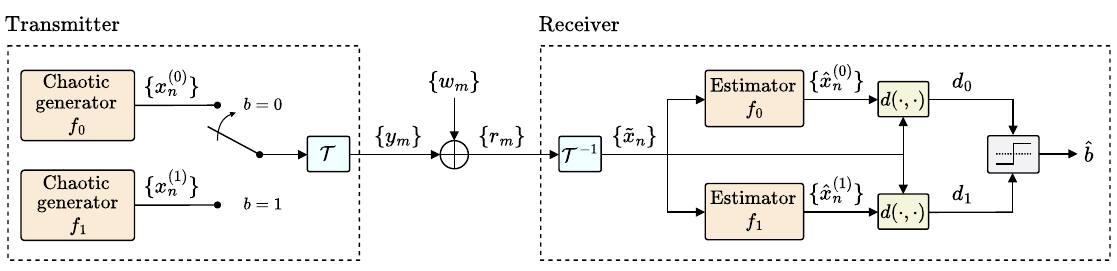}
  \caption{Transmitter and receiver designs for the predictive chaotic communication scheme.}
  \label{fig: communication_scheme}
\end{figure*}

Figure \ref{fig: communication_scheme} shows a block diagram of the proposed communication scheme. The system model is general enough to encompass any chaotic estimator and any distance metric.

\subsection{Noisy channel}
In the following we will show that the proposed estimation in \eqref{eq:prediction} for predictive chaotic demodulation is not optimal in an AWGN channel. Specifically, classical correlation-based methods (e.g., CSK) process the data in a linear fashion, which does not alter the noise distribution. Nonetheless, the predictive demodulation introduced in \ref{sec: noiseless_channel} is nonlinear, because the samples are predicted using the nonlinear chaotic map $f_i(\cdot)$. These transformations change the noise distribution and do not minimize the MSE in \eqref{eq: distance_metric}.
For instance, consider the following quadratic and trigonometric maps
\begin{equation}
f(x_{n}) =
\begin{cases}
a - x_{n}^2 &
\text{if $b=0$}\\
A\cos\left(x_{n}+\phi\right) &
\text{if $b=1$}
\end{cases}
\label{eq: chaotic_maps}
\end{equation}
where $a, A$ and $\phi$ are parameters that characterize the chaotic maps. To study the effect of the noise in the prediction, we assume no modulation (i.e., $y_m=x_n$). Then, estimating \eqref{eq:prediction} in the presence of noise turns into 
\begin{equation}
\hat{x}_{n+1}^{(i)}=f_i(x_n+w_n) =
\begin{cases}
x_{n+1} -2x_nw_n - w_n^2 &
\text{if $i=0$}\\
A\cos\left(x_n+w_n+\phi\right) &
\text{if $i=1$}
\end{cases}
\label{eq: noisy_chaotic_maps}
\end{equation}

When the noise is small with respect to the information signal, \eqref{eq: noisy_chaotic_maps} can be approximated as
\begin{equation}
\hat{x}_{n+1}^{(i)}\approx
\begin{cases}
x_{n+1} -2x_nw_n &
\text{if $i=0$}\\
x_{n+1}-Aw_n\sin\left(x_n+\phi\right) &
\text{if $i=1$}
\end{cases}
\label{eq: low_noise_maps}
\end{equation}
which indicate two different behaviours of the noise: in the quadratic map, the noise becomes correlated with the chaotic signal; conversely, in the trigonometric map, the noise exhibits the characteristic effect seen in angular modulations, in which small variations of
received signal power cause sizable changes in the output signal, causing a threshold effect under which the system is impractical \cite{carlson}.

More generally, the predictions performed under white noise incur in biased estimations. In the case of the quadratic map, the best estimate of the next chaotic sample is
\begin{align}
\mathbb{E}\{x_{n+1}^{(0)}\}&=
\frac{1}{\sqrt{2\pi\sigma^2}}
\int_{-\infty}^{\infty} \left(
a-(x_n+w)^2
\right)
e^{-\frac{w^2}{2\sigma^2}}
dw\nonumber\\
&=a-x_n^2-\sigma^2,
\end{align}
where $\mathbb{E}$ is the expected value over the noise distribution. Thus, the best estimate for the next sample is not $f_0(x_n)$ but $f_0(x_n)-\sigma^2$. The average MSE in \eqref{eq: distance_metric} yields an improvement of $2\sigma^4$ with the unbiased estimator \cite{Schreiber1996_2percentnoise}. For the trigonometric map, the mean estimation is $f_1(x_n)\exp\left({-\sigma^2/2}\right)$, and the reduction in MSE cannot be found in a closed-form expression.

Making $P$ predictions more than one step ahead provides a set of estimates $\{\hat{x}_{n,p}^{(i)}\}$ that reduces the noise effect by a factor of $P$ when compared with the received signal in \eqref{eq: distance_metric}. However, this requires averaging over all possible perturbations at the intermediate time steps. Under realistic
conditions the noise is unknown with the accuracy necessary for this strategy. 

In conclusion, the estimation has to consider the noise distribution as well as the nature of the chaotic map. In practice, the latter has a non-negligible effect in the performance of chaotic communications, even in correlation-based approaches \cite{tse2003chaos}.
The effect of additive noise in the proposed scheme is more severe than in correlation-based schemes due to the nonlinear characteristics of the method. In this respect, the flat spectral density of chaotic signals implies that samples with small amplitude are equally likely to occur. Then, estimating the next sample using these noisy samples is equivalent to random guessing, as the sample is almost noise. These wrong estimations pay a high cost in \eqref{eq: distance_metric}.

In the following section we introduce a modulation that is more robust to noise, preserves the random-like nature of chaotic signals and is suited for predictive strategies.

\section{Frequency modulations for chaotic signals}
Using the chaotic signal as a waveform is impractical for several reasons: first, the transmission bandwidth is theoretically infinite and depends on the chaotic map; second, due to the non-periodic behavior, the energy of the sequence depends on the map and the initial condition $x_0$, which can only be estimated. Since the performance depends on the energy, there are no guarantees on the system performance; and third, noise highly degrades the performance of predictive demodulation, as shown in the previous section.

To overcome these issues, we propose to modulate the chaotic signal in the frequency domain. The benefits of frequency modulations for chaotic communications have already been exploited in the past. Particularly, the authors in \cite{kolumban1998fm} propose to feed an analog frequency modulator (FM) with a chaotic signal, whose output will be used for differential CSK (DCSK). This provides a wideband signal with constant energy per bit, that is also more robust to noise.

Given the widespread of digital modulations, we propose a method that relies on digital frequency modulations. First, we process the chaotic signal with an $M$-level uniform quantizer $\mathcal{Q}:\mathbb{R}\rightarrow[0,M-1]$:
\begin{equation}
\bar{x}_n=\mathcal{Q}(x_n)
\in[0,M-1]~~
n=0,\dots,N-1,
\end{equation}
In general, there is no reason to assume a nonuniform quantizer, due to the white spectral nature of chaotic signals.
Each quantized sample is fed to an $M$-ary FSK modulator:
\begin{equation}
y_m=\sqrt{\frac{2}{N}}\cos\left(
\frac{\pi(2\bar{x}_n+1)}{2M}m
\right),
~~
m=0,\dots,M-1,
\label{eq:FSK}
\end{equation}
where $m$ is the time index of the modulated signal and $M$ is the total number of transmitted samples. Note that the transmitted signal is still chaotic, as it is a deterministic transformation over a chaotic signal. Finally, the receiver recovers the quantized measurement with a bank of filters at the $M$ frequencies.

This scheme overcomes the impairments of transmitting the chaotic signal over the wireless channel: First, \eqref{eq:FSK} is an angular modulation, in which the reconstruction is noiseless as long as the systems works above a threshold signal-to-noise ratio (SNR). The error probability in demodulation (i.e., $\tilde{x}_n=\bar{x}_n$) can be approximated as \cite{martinez2023feel}
\begin{equation}
P_e \approx (M-1)
Q\left(\sqrt{\frac{E_b}{MN_o}}\right),
\label{eq:error_probability_FSK}
\end{equation}
where $Q$ is the $Q$-function and $E_b$ is the energy per bit associated to the transmitted signal. Note that the length of the transmitted waveform reduces the SNR because, for a fixed chaotic sequence of length $N$ and associated energy $E_b$, the FSK modulation increases the length of the transmitted waveform by a factor of $M$. This presents a trade-off between the quantization noise and the error probability: The larger the $M$, the finer the quantization and the smaller the quantization noise, although the longer the time signal and the smaller the energy per transmitted sample. Nonetheless, in \ref{sec:results} we empirically prove that a small $M$ provides the best performance. Moreover, note that the $M$ FSK signals can be further multiplexed in the frequency domain (e.g., with chirp signals \cite{martinez2023feel}), which does not increase the delay of the system.

Second, the transmitted power is constant and independent of the chaotic signal, and the signal $\{y_m\}$ is more robust to noise than $\{x_n\}$ due to the frequency nature of the signal \cite{carlson}. And third, the bandwidth is now finite. 
Assuming a sampling frequency of $f_s = M/T$, where $T$ is
the duration of the chaotic sample $\bar{x}_n$, it results in a maximum bandwidth of $B=M/2T$.

\section{Results}
\label{sec:results}
Due to the inherent nonlinear structure of the predictive scheme, the BER cannot be obtained in a closed-form expression. Furthermore, as in conventional chaotic schemes, the performance also depends on the chaotic maps. Thus, in the following we provide experimental simulations on the BER of the predictive demodulation for chaotic communications.

We use the chaotic maps in \eqref{eq: chaotic_maps} with $a=1.6$, $A=2.2$ and $\phi=47\pi/64$.
We generate $K=10^5$ bits, and each bit $b_k$ determines $f_0(\cdot)$ or $f_1(\cdot)$ to generate a sequence $\{x_n\}$. The initial condition $x_0$ for each chaotic sequence is generated randomly in the $[-1,1]$ range. Each chaotic sequence $\{x_n\}$ is normalized to satisfy $|x_n|<1, \forall n$, and the mean is subtracted. The chaotic sequence is quantized to $M$ levels in the $[-1,1]$ range and fed to an $M$-ary FSK modulator. For a given $E_b$, each modulated signal is transmitted with an energy of $E_b/M$ over an AWGN channel. The receiver demodulates the signal and uses the chaotic sample to estimate the succeeding.

\begin{figure}[t]
\centering
\includegraphics[width=\columnwidth]{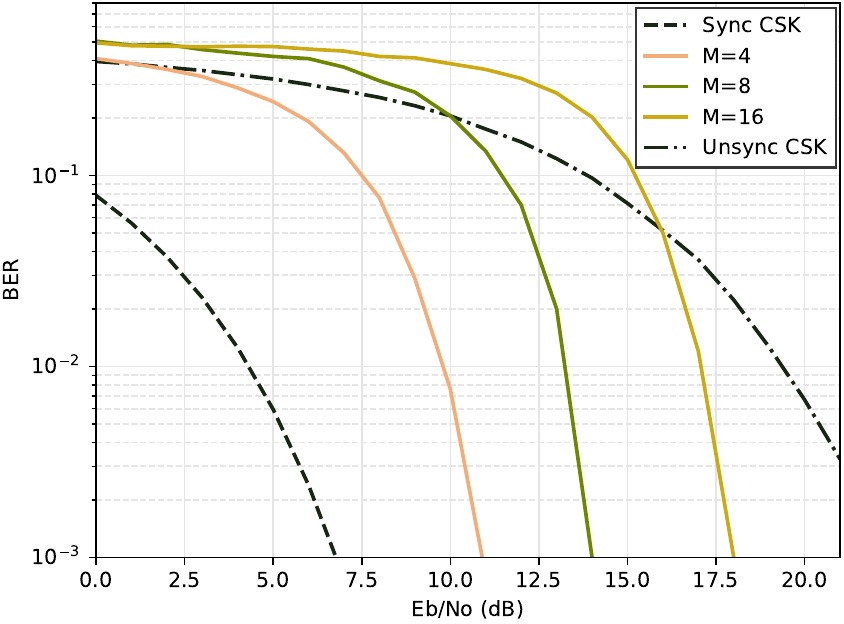}
\caption{BER of different chaotic communication schemes for $N=128$. As benchmarks we show the synchronized CSK (lower bound), which highly degrades without perfectly synchronized replicas. The predictive scheme with frequency modulations is presented for $M=\{4,8,16\}$.}
\label{fig:sim_ber_chaos}
\vspace{-14 pt}
\end{figure}

Figure \ref{fig:sim_ber_chaos} depicts
the BER with respect to the $E_b/N_o$ ratio for different chaotic communication schemes and $N=128$. As a benchmark, the synchronized CSK scheme corresponds to a classical correlation-based demodulator with perfect knowledge of the transmitted chaotic waveforms. This scheme achieves the performance of a BPSK, i.e., $Q\left(\sqrt{E_b/N_o}\right)$ \cite{lawrence2003exactber}, which is used as a lower bound on the BER. The unsynchronized CSK represents the same system, without perfect knowledge at the receiver. To generate the unsynchronized replicas at the receiver, we use the maps in \eqref{eq: chaotic_maps} with a small perturbation in the initial condition, i.e., $x_0+\varepsilon$, and $\varepsilon=10^{-8}$. As expected, the performance highly degrades because the replicas available at the receiver become uncorrelated with the transmitted ones, even at high SNR. This proves that correlation-based schemes cannot be used without perfectly synchronized chaotic waveforms.

The predictive scheme is tested for different $M$-ary FSK modulations, namely, $M=\{4,8,16\}$. At high SNR, prediction schemes outperform unsynchronized correlation schemes. At lower SNR, the lower the order of the modulation, the better the performance. This is because the effect of the energy reduction when increasing $M$ is more detrimental than the increase in quantization noise.
For $M=4$, the prediction scheme outperforms the unsynchronized CSK in all SNR regimes and delivers a small increase in the signal length. Finally, the performance of the proposed scheme can be improved by increasing the length of the chaotic sequence, $N$, which approaches the lower bound.

\section{Conclusion}
In this paper we present the first modulation for chaotic communications that exploits the deterministic nature of these signals. Unlike CSK, the proposed scheme does not require synchronized copies of the transmitted waveforms at the receiver side. To make the signal more robust against the noise, we modulate the chaotic samples with $M$-ary FSK. We empirically show that the proposed scheme outperforms CSK when perfect replicas are not available at the receiver. The best performance is ahieved for low-order modulations (e.g., $M=4$), which also incur in a very small increase in delay.

\bibliographystyle{IEEEbib}
\bibliography{refs}

\end{document}